\documentclass[aps,pre,twocolumn,showpacs,floatfix,longbibliography,superscriptaddress,noeprint]{revtex4-2}
\usepackage{graphicx,amsfonts,amssymb,amsmath}
\usepackage{textcomp}
\usepackage{esint}
\usepackage[colorlinks,linktocpage,bookmarks=false,citecolor=blue,linkcolor=red,urlcolor=blue]{hyperref}
\usepackage[utf8]{inputenc}
\usepackage[dvipsnames]{xcolor}
\usepackage[english]{babel}
\usepackage{microtype}
\usepackage{afterpage}
\usepackage{bbold}
\usepackage{soul}
\usepackage{siunitx}
\usepackage[capitalize]{cleveref}
\usepackage[normalem]{ulem}

\input cyracc.def

\allowdisplaybreaks

\begin{document}
	

\def\rhoeq{\hat\rho_{\rm eq}}


\newcommand{\beq}{\begin{equation}}
\newcommand{\eeq}{\end{equation}}
\newcommand{\bfig}{\begin{figure}}
\newcommand{\efig}{\end{figure}}
\newcommand{\bline}{\begin{multline}}
\newcommand{\eline}{\end{multline}}
\newcommand{\bremark}{\begin{quotation} \noindent \small }
\newcommand{\eremark}{\end{quotation}}
\newcommand{\llbrace}{\left\lbrace}  
\newcommand{\rrbrace}{\right\rbrace}
\newcommand{\lbraket}{\left[}
\newcommand{\rbraket}{\right]}
\newcommand{\llangle}{\left\langle}
\newcommand{\rrangle}{\right\rangle} 

\newcommand{\Tr}{{\rm Tr}} 
\newcommand{\tr}{{\rm tr}} 
\newcommand{\sgn}{\,{\rm sgn}} 
\newcommand{\mean}[1]{\langle #1 \rangle}
\newcommand{\commu}[2]{[#1,#2]} 
\newcommand{\bra}[1]{\langle#1|}
\newcommand{\ket}[1]{|#1\rangle}
\newcommand{\braket}[2]{\langle #1|#2\rangle}
\newcommand{\ketbra}[2]{|#1\rangle\langle#2|}
\newcommand{\dbraket}[3]{\langle #1|#2|#3\rangle}
\newcommand{\tens}[1]{\overleftrightarrow{#1}}  
\newcommand{\vac}{|{\rm vac}\rangle} 
\newcommand{\bravac}{\langle{\rm vac}|}
\newcommand{\const}{{\rm const}} 
\newcommand{\unif}{{\rm unif.}} 
\newcommand{\atanh}{\,{\rm atanh}}
\newcommand{\cotanh}{\,{\rm cotanh}}

\newcommand{\ie}{i.e.\xspace}
\newcommand{\iet}{i.e.}
\newcommand{\eg}{e.g.\xspace}
\newcommand{\cc}{{\rm c.c.}} 
\newcommand{\hc}{{\rm h.c.}} 
\newcommand{\etal}{{\it et al. }}
\newcommand\eme{$^{\mbox{\footnotesize ème}}$\xspace}

\newcommand{\jhatbf}{\hat {\textbf \jold}} 
\newcommand{\Jhatbf}{\hat {\textbf \J}} 
\newcommand{\jhat}{\hat {\jmath}} 
\newcommand{\Jhat}{\hat {J}} 
\newcommand{\jbf}{\textbf j}
\newcommand{\Jbf}{\textbf J}

\def\chibf{\boldsymbol{\chi}}
\def\down{\downarrow}
\def\eps{\epsilon}
\def\gam{\gamma} 
\def\alphabf{\boldsymbol{\alpha}}
\def\phibf{\boldsymbol{\phi}}
\def\varphibf{\boldsymbol{\varphi}}
\def\varphibfs{\boldsymbol{\varphi}_<}
\def\varphibfl{\boldsymbol{\varphi}_>}
\def\varphis{\varphi_{<}}
\def\varphil{\varphi_{>}}
\def\psibf{\boldsymbol{\psi}}
\def\thetabf{\boldsymbol{\theta}}
\def\Ome{\Omega}
\def\omeD{{\omega_D}} 
\def\bfOme{\boldsymbol{\Omega}} 
\def\Omebf{\boldsymbol{\Omega}} 
\def\lamb{\lambda}
\def\Lamb{\Lambda}
\def\sig{\sigma}
\def\Sig{\Sigma}
\def\sigp{{\sigma'}} 
\def\bfsig{\boldsymbol{\sigma}} 
\def\sigbf{\boldsymbol{\sigma}} 
\def\bfSig{\boldsymbol{\Sigma}} 
\def\The{\Theta} 
\def\up{\uparrow}

\def\epsk{\epsilon_{\bf k}} 
\def\xik{\xi_{\bf k}} 
\def\txik{\tilde\xi_{\bf k}} 
\def\xip{\xi_{\bf p}} 
\def\xiq{\xi_{\bf q}} 
\def\xikq{\xi_{{\bf k}+{\bf q}}} 
\def\Ek{E_{\bf k}} 
\def\Ep{E_{\bf p}}
\def\Eq{E_{\bf q}}
\def\Heff{\hat H_{\rm eff}}
\def\Hem{\hat H_{\rm em}}
\def\Hint{\hat H_{\rm int}}
\def\Hloc{\hat H_{\rm loc}}
\def\HMF{\hat H_{\rm MF}}
\def\HLL{\hat H_{\rm LL}}
\def\Sem{S_{\rm em}}
\def\SMF{S_{\rm MF}} 
\def\SHF{S_{\rm HF}} 
\def\SRPA{S_{\rm RPA}} 
\def\Sint{S_{\rm int}} 
\def\Sloc{S_{\rm loc}}
\def\TN{T_{\rm N}} 
\def\TNHF{T^{\rm HF}_{\rm N}} 
\def\Zloc{Z_{\rm loc}} 
\def\ZMF{Z_{\rm MF}} 
\def\ZHF{Z_{\rm HF}} 
\def\ZRPA{Z_{\rm RPA}} 
\def\RPA{{\rm RPA}}
\def\loc{{\rm loc}} 
\def\pp{{\rm pp}}
\def\ph{{\rm ph}} 
\def\ch{{\rm ch}}
\def\sp{{\rm sp}} 
\def\qtf{q_{\rm TF}}
\def\epstf{\eps^{}_{\rm TF}} 
\def\epsrpa{\eps^{}_{\rm RPA}} 
\def\chinnzpp{\chi_{nn}^{0}{}\!\!\!''}

\def\half{\frac{1}{2}}
\def\dhalf{\dfrac{1}{2}}
\def\third{\frac{1}{3}} 
\def\quarter{\frac{1}{4}}

\def\qr{{\bf q}\cdot{\bf r}}
\def\wt{\omega t} 

\def\a{{\bf a}}
\def\b{{\bf b}}
\newcommand{\cv}{{\bf c}} 
\def\e{{\bf e}}
\def\f{{\bf f}}
\def\g{{\bf g}}
\def\h{{\bf h}}
\def\jold{\char"11}
\def\j{{\bf j}}
\def\k{{\bf k}}
\def\l{{\bf l}}
\def\ellbf{\bm{\ell}} 
\def\m{{\bf m}}
\def\n{{\bf n}} 
\def\p{{\bf p}} 
\def\q{{\bf q}}
\def\r{{\bf r}}
\def\t{{\bf t}}
\def\u{{\bf u}}
\newcommand{\vv}{{\bf v}}
\def\x{{\bf x}}
\def\y{{\bf y}} 
\def\z{{\bf z}} 
\def\A{{\bf A}}
\def\B{{\bf B}}
\def\D{{\bf D}} 
\def\E{{\bf E}} 
\def\F{{\bf F}} 
\def\H{{\bf H}}  
\def\J{{\bf J}}
\def\K{{\bf K}} 

\def\G{{\bf G}}
\def\L{{\bf L}}
\def\M{{\bf M}}  
\def\O{{\bf O}} 
\def\P{{\bf P}} 
\def\Q{{\bf Q}} 
\def\R{{\bf R}}
\def\S{{\bf S}}
\def\U{{\bf U}} 
\def\V{{\bf V}} 
\def\X{{\bf X}} 
\def\Y{{\bf Y}} 
\def\epsbf{\boldsymbol{\epsilon}}
\def\betabf{\boldsymbol{\beta}}
\def\deltabf{\boldsymbol{\delta}}
\def\mubf{\boldsymbol{\mu}}
\def\nablabf{\boldsymbol{\nabla}}
\def\rhobf{\boldsymbol{\rho}}
\def\sigmabf{\boldsymbol{\sigma}} 
\def\Pibf{\boldsymbol{\Pi}}
\def\pibf{\boldsymbol{\pi}}

\def\para{\parallel}
\def\kpara{{k_\parallel}}
\def\kperp{{k_\perp}} 
\def\kperpp{{k_\perp'}} 
\def\qperp{{q_\perp}} 
\def\tperp{{t_\perp}} 

\def\w{\omega}
\def\wn{\omega_n}
\def\wm{\omega_m}
\def\wnu{\omega_\nu}
\def\wp{\omega_p} 
\def\dmu{{\partial_\mu}}
\def\dnu{{\partial_\nu}}
\def\dl{{\partial_l}}  
\def\dt{\partial_t} 
\def\tdt{\tilde\partial_t}
\def\dk{\partial_k}
\def\tdk{\tilde\partial_k}
\def\dx{\partial_x}
\def\dy{\partial_y} 
\def\dtau{{\partial_\tau}}  
\def\det{{\rm det}} 
\def\Pf{{\rm Pf}}
\def\diag{{\rm diag}}

\def\dsum{\displaystyle \sum}
\def\dint{\displaystyle \int} 
\def\intt{\int_{-\infty}^\infty dt} 
\def\inttp{\int_{-\infty}^\infty dt'} 
\def\intk{\int_{\bf k}} 
\def\intkd{\int \frac{d^dk}{(2\pi)^d}}
\def\intq{\int_{\bf q}} 
\def\intr{\int d^dr}  
\def\dintr{\displaystyle \int d^dr} 
\def\intrp{\int d^dr'}
\def\dinttau{\displaystyle \int_0^\beta d\tau}
\def\dinttaup{\displaystyle \int_0^\beta d\tau'}
\def\inttau{\int_0^\beta d\tau}
\def\inttaup{\int_0^\beta d\tau'}
\def\intx{\int d^{d+1}x} 
\def\inttaur{\int_0^\beta d\tau \int d^dr}
\def\intinf{\int_{-\infty}^\infty}
\def\dinttaur{\displaystyle \int_0^\beta d\tau \int d^dr}
\def\dintinf{\displaystyle \int_{-\infty}^\infty}
\def\intw{\int_{-\infty}^\infty \frac{d\w}{2\pi}}
\def\sumr{\sum_{\bf r}} 

\def\calA{{\cal A}}
\def\calAbf{\bm{{\cal A}}}
\def\calB{{\cal B}} 
\def\calC{{\cal C}} 
\def\dt{\partial_t}
\def\calD{{\cal D}}
\def\calE{{\cal E}}
\def\calF{{\cal F}} 
\def\calFbf{\bm{{\cal F}}}
\def\calG{{\cal G}}
\def\calH{{\cal H}}
\def\calI{{\cal I}}
\def\calJ{{\cal J}}
\def\calK{{\cal K}}
\def\calL{{\cal L}} 
\def\calM{{\cal M}} 
\def\calN{{\cal N}}
\def\calO{{\cal O}}
\def\calP{{\cal P}}  
\def\calR{{\cal R}} 
\def\calS{{\cal S}}
\def\calT{{\cal T}}
\def\calU{{\cal U}}
\def\calV{{\cal V}}
\def\calX{{\cal X}} 
\def\calY{{\cal Y}} 
\def\calZ{{\cal Z}} 

\def\calbfB{{\bf \cal B}}
\def\calbfF{{\bf \cal F}}

\def\tT{{\tilde T}}
\def\talpha{{\tilde\alpha}}
\def\tbeta{{\tilde\beta}}
\def\tchi{{\tilde\chi}}
\def\tdelta{{\tilde\delta}}
\def\tDelta{{\tilde\Delta}}
\def\teta{{\tilde\eta}} 
\def\tlamb{{\tilde\lambda}}
\def\tmu{{\tilde\mu}}
\def\tphibf{{\tilde\phibf}}
\def\trho{{\tilde\rho}}
\def\tvarphibf{{\tilde\varphibf}} 
\def\tw{{\tilde\omega}}
\def\twn{{\tilde\omega_n}}
\def\twnu{{\tilde\omega_\nu}}

\def\asinh{{\rm asinh}} 
\def\Tbkt{T_{\rm BKT}}
	
	\graphicspath{{./figures/}}
	
\title{Operator product expansion coefficients from the nonperturbative functional renormalization group}
	
	\author{Félix Rose}
\affiliation{Max Planck Institute of Quantum Optics, Hans-Kopfermann-Stra{\ss}e 1, 85748 Garching, Germany} 
\affiliation{Munich Center for Quantum Science and Technology (MCQST),
	Schellingstra{\ss}e 4, 80799 Munich, Germany
}
	\author{Carlo Pagani}
	\affiliation{Université Grenoble Alpes, Centre National de la Recherche Scientifique, Laboratoire de Physique et Modélisation des Milieux
	Condensés, 38000 Grenoble, France}
    \affiliation{Institute of Physics, Johannes Gutenberg University,
Staudingerweg 7, 55099 Mainz, Germany}
	\author{Nicolas Dupuis}
	\affiliation{Sorbonne Universit\'e, CNRS, Laboratoire de Physique Th\'eorique de la Mati\`ere Condens\'ee, LPTMC, F-75005 Paris, France}
	
	\date{April 1, 2022} 
	
\begin{abstract}

Using the nonperturbative functional renormalization group (FRG) within the Blaizot-M\'endez-Galain-Wschebor approximation, we compute the operator product expansion (OPE) coefficient $c_{112}$ associated with the operators $\calO_1\sim\varphi$ and $\calO_2\sim\varphi^2$ in the three-dimensional $\mathrm{O}(N)$ universality class and in the Ising universality class ($N=1$) in dimensions $2 \leq d \leq 4$. When available, exact results and estimates from the conformal bootstrap and Monte Carlo simulations compare extremely well to our results, while FRG is able to provide values across the whole range of $d$ and $N$ considered.
\end{abstract}

\maketitle
	
\tableofcontents

\section{Introduction} 

The nonperturbative functional renormalization group (FRG) provides us with a versatile technique to study strongly correlated systems. It has been used in many models of quantum and statistical field theory ranging from statistical physics and condensed matter to high-energy physics and quantum gravity~\cite{Berges02,Delamotte12,Dupuis_review}. Besides the interest in models where perturbative approaches or numerical methods are difficult for various reasons, there is an ongoing effort to characterize and quantify the efficiency and the accuracy of the FRG approach by considering well-known models of statistical physics. It is now proven that the FRG yields very accurate values of the critical exponents associated with the Wilson-Fisher fixed point of $\mathrm{O}(N)$ models~\cite{Balog19,DePolsi20}, comparable with the best estimates from field-theoretical perturbative RG~\cite{Guida98,Kompaniets17}, Monte Carlo simulations~\cite{Hasenbusch10,Campostrini06,Campostrini02,Hasenbusch19,Clisby16,Clisby17} or conformal bootstrap~\cite{Kos16,SimmonsDuffin17,Echeverri16,Chester20}. The FRG also allows the computation of universal quantities defined away from the critical point, such as universal scaling functions~\cite{Berges96,Rancon13a,Rancon13b,Rancon16} or universal amplitude ratios~\cite{DePolsi2021a}, again in remarkable agreement with Monte Carlo simulations when available.

On the other hand, the operator product expansion (OPE) has received little attention in the framework of the FRG until recently~\cite{Hughes:1988cp,Keller:1991bz,Keller:1992by,Hollands:2011gf,Holland:2014ifa,Pagani:2017tdr,Pagani20}.
Wilson and Kadanoff suggested independently that in a  quantum field theory 
the product of two operators in the short distance limit
is equivalent to
an infinite sum of operators multiplied by possibly singular functions
when inserted in any correlation function~\cite{Wilson1965a,Wilson69a,Kadanoff1969a,Wilson1972a}. 
The validity of the OPE has been proven to all orders in pertubation theory~\cite{Zimmermann:1972tv}
and 
can be established in full generality in the case of conformal field theories~\cite{Rychkov2017a}.
Indeed, the OPE has been fundamental in the study of conformal field theories in two
and higher dimensions \cite{Belavin:1984vu,Poland2019a}.
In this context, the conformal bootstrap program~\cite{Ferrara1973a,Polyakov1974a,Rattazzi:2008pe,Poland2019a} 
has lead to a large number of precise results. 
The OPE has been instrumental as well in studies regarding quantum chromodynamics \cite{Novikov:1977dq}
and condensed matter, where it has been used to derive the thermodynamic properties of  quantum gases~\cite{Olshanii2003a,Barth2011a}.

Despite the fact that both the FRG formalism and the OPE offer non-perturbative approaches to
quantum field theory,
it is not yet clear to what extent these two aspects can be usefully combined to extract
information regarding the non-perturbative regime of a field theory.

From the perspective of perturbation theory,
the FRG provides a useful framework that allows one to prove the existence of the OPE
perturbatively~\cite{Hughes:1988cp,Keller:1991bz,Keller:1992by,Hollands:2011gf,Holland:2014ifa}.
Moreover, by following the proposal of Cardy relating the OPE coefficients to the second order
terms in the expansion of the beta functions around a fixed point~\cite{Cardy:1996xt},
the standard perturbative renormalization group has been used to derive
certain OPE coefficients within the $\epsilon$ expansion~\cite{Codello:2017hhh};
we refer to~\cite{Pagani:2017gnd} for an FRG perspective on these issues
based on a geometric approach to theory space.

In principle, one may reconstruct from the FRG the full operator product and express the latter
as an OPE~\cite{Pagani:2017tdr,Pagani20}.
However, this may be rather cumbersome in practice.
For a conformally invariant fixed point theory~\footnote{Note that conformal invariance, and in particular its relation to scale invariance, has been discussed in a number of recent works based on the FRG, see Refs.~\cite{Delamotte16a,Delamotte18,DePolsi18,DePolsi19,Sonoda:2017zgl}.},
a further possibility explored in~\cite{Pagani20} consists in extracting the OPE coefficient
 from  three-point functions.
It has been shown that within this approach it is possible to calculate the 
OPE coefficients in the epsilon expansion.

The main quantities of interest in the FRG are the effective action, defined as the Legendre transform of the free energy, and the one-particle irreducible (1PI) vertices. Taking the Wilson-Fisher fixed point of the $\mathrm{O}(N)$ model as an example, we show how the OPE coefficient $c_{112}$ associated with the operators $\calO_1\sim\varphi$ and $\calO_2\sim\varphi^2$ can be deduced from a small number of low-order 1PI vertices. 
One difficulty in the computation of  OPE coefficients is that the latter are 
determined 
by the full momentum dependence of the vertices in the critical regime.  
For this reason, one has to go beyond the derivative expansion in order to accurately determine the OPE coefficients. The latter can be computed
in the so-called Blaizot-M\'endez-Galain-Wschebor (BMW) approximation that enables the determination of the momentum dependence of the correlation functions~\cite{Blaizot06,Benitez09,Benitez12}. This approximation scheme has been used in the past to obtain the spectral function of the ``Higgs'' amplitude mode in the $(2+1)$-dimensional $\mathrm{O}(N)$ model~\cite{Rose15} providing an estimate of the Higgs mass that has been confirmed by subsequent numerical simulations of lattice models~\cite{Lohofer15,Nishiyama15,Nishiyama16}.

The outline of the paper is as follows. In \cref{sec_OPEcoef} we recall the relation between the OPE coefficients and the two- and three-point functions in momentum space, focusing on the coefficient $c_{112}$ in the $d$-dimensional $\mathrm{O}(N)$ model. We then show how to relate $c_{112}$ to the 1PI vertices. Finally we briefly describe the nonperturbative FRG formalism and the BMW approximation. In \cref{sec_results} the results obtained from a numerical solution of the flow equations are discussed for the three-dimensional $\mathrm{O}(N)$ model  and the Ising university class ($N=1$) in dimensions $2\leq d\leq 4$, and compared with exact values in some particular cases and estimates from conformal bootstrap and Monte Carlo as well as $\epsilon$ and large-$N$ expansions.

\section{OPE coefficients in the effective action formalism}
\label{sec_OPEcoef} 

\subsection{Correlation functions in momentum space} 

We consider a critical, conformally invariant, theory. For fields $\calO_a(x)$ (be them composite or not) with scaling dimensions $\Delta_a$, the two- and three-point correlation functions are given by 
\beq 
\mean{\calO_a(x) \calO_a(y)} = \frac{1}{|x-y|^{2\Delta_a}} 
\label{2ptcf}
\eeq  
and
\begin{multline} 
\mean{\calO_a(x_1) \calO_b(x_2) \calO_c(x_3)} = \\  \frac{c_{abc}}{x_{12}^{\Delta_a+\Delta_b-\Delta_c} x_{23}^{\Delta_b+\Delta_c-\Delta_a} x_{13}^{\Delta_a+\Delta_c-\Delta_b} }
\label{3ptcf}
\end{multline}
where $x_{12}=|x_1-x_2|$, etc. \Cref{2ptcf} assumes a proper normalization of the fields and the coefficient $c_{abc}$ in~\labelcref{3ptcf} can be identified with the OPE coefficient~\cite{Francesco1997a}. Since in practice we shall work in momentum space, it is convenient to consider the Fourier transformed correlation functions. For the two-point one, 
\begin{align}
\mean{\calO_a(p) \calO_a(-p)} = \int_x \frac{e^{-ipx}}{|x|^{2\Delta_a}} 
= \frac{A_d(\Delta_a)}{|p|^{d-2\Delta_a}} ,
\label{ope1} 
\end{align} 
where 
\beq 
A_d(\Delta) = 4^{d/2-\Delta} \pi^{d/2} \frac{\Gamma(d/2-\Delta)}{\Gamma(\Delta)} 
\eeq 
with $\Gamma(x)$ the gamma function and $d$ the dimension. The Fourier transform $\mean{\calO_a(p_1) \calO_b(p_2) \calO_c(p_3)}$ is given by a complicated expression but it is sufficient to consider the limit $|p_1|\gg|p_2|$, where  
\begin{multline}
\mean{\calO_a(p_1) \calO_b(p_2) \calO_c(-p_1-p_2)} \simeq \\ \frac{c_{abc} A_d((\Delta_a-\Delta_b+\Delta_c)/2) A_d(\Delta_b)}{|p_1|^{d-\Delta_a+\Delta_b-\Delta_c} |p_2|^{d-2\Delta_b}} ,
\label{ope2} 
\end{multline}
to extract the coefficient $c_{abc}$~\cite{Pagani20}. 
\Cref{ope2} entails that the OPE coefficient $c_{abc}$
can be deduced from the three-point function~\labelcref{ope2}
once the fields have been properly normalized in order to satisfy~\labelcref{ope1}.

\subsection{$\mathrm{O}(N)$ model and Wilson-Fisher fixed point} 

In the following we consider the $\mathrm{O}(N)$ model in $d$ dimensions defined by the action
\beq 
S[\varphi] =\int_x \llbrace \half (\dmu\varphi)^2 + \frac{r_0}{2} \varphi^2 + \frac{u_0}{4!N} {(\varphi^2)}^2 \rrbrace  
\label{action} 
\eeq 
and regularized by a UV momentum cutoff $\Lambda$. $\varphi=(\varphi_1,\cdots,\varphi_N)$ is a $N$-component field. The model can be tuned to its critical point by varying $r_0$. The correlation functions are then scale and conformal invariant in the momentum range $|p|\ll p_G$ where $p_G\sim u_0^{1/(d-4)}$ is the Ginzburg scale. In the following we shall only be interested in the critical point and the scaling limit $|p|\ll p_G$; we refer to \cite{Dupuis2011} for an overview of the various regimes of the
$O\left(N\right)$ model.

We focus on the operators 
\beq 
\begin{split} 
\calO_1(x) &= \calN_1 \varphi_{i}(x) , \\
\calO_2(x) &= \calN_2 \frac{\varphi(x)^2}{2} 
\end{split}
\label{def:composite-op-initial-cond}
\eeq 
(the index $i$ is arbitrary) and the OPE coefficient $c_{112}$.  Note that the correlation functions $\mean{\calO_1 \calO_1}$ and $\mean{\calO_1\calO_1 \calO_2}$ are independent of $i$ at criticality and in the whole disordered phase. $\calN_1$ and $\calN_2$ are normalization constants that ensure that $\mean{\calO_1(p)\calO_1(-p)}$ and $\mean{\calO_2(p)\calO_2(-p)}$ are given by~\labelcref{ope1}  in the scaling limit $|p|\ll p_G$. Even though $\calO_2$ is not, strictly speaking, a scaling operator, it can be expressed as a linear combination of scaling operators, among which that associated with $\Delta_2$. In the scaling limit, corrections due to higher scaling dimension operators are suppressed and we neglect them; a more detailed explanation regarding this point is provided at the end of \cref{sec:frgbmw}.
The scaling dimension 
\beq 
\Delta_1=[\varphi_i]=\frac{d-2+\eta}{2}
\eeq 
is related to the anomalous dimension $\eta$ while 
\beq 
\Delta_2=[\varphi^2]=d-\frac{1}{\nu}
\eeq 
where $\nu$ is the correlation-length exponent.

To deal with the composite field $\calO_2$, in addition to the linear source $J$ we introduce  in the partition function a source $h$ coupled to $\varphi^2$,
\beq 
\calZ[J,h] = \int\calD[\varphi] \, e^{-S[\varphi] + \int_x (J\varphi + h\varphi^2) } .
\eeq 
The correlation functions of interest, besides the propagator $G(x-y)=\mean{\varphi_{i}(x)\varphi_{i}(y)}_c$ for $h=0$ ($\mean{\cdots}_c$ stands for the connected correlation function), are the scalar susceptibility 
\beq 
\chi_s(x-y) = \mean{\varphi(x)^2 \varphi(y)^2}_c = \frac{\delta^2 \ln \calZ[J,h]}{\delta h(x) \delta h(y)} \biggr|_{J=h=0}
\eeq
and the three-point function 
\begin{align} 	
\chi(x,y,z) &= \mean{\varphi_{i}(x) \varphi_{i}(y) \varphi(z)^2}_c \nonumber\\ 
&= \frac{\delta^3 \ln \calZ[J,h]}{\delta J_{i}(x) \delta J_{i}(y) \delta h(z)} \biggr|_{J=h=0} .
\end{align} 
Here and in the following, there is no implicit summation over the index $i$.
The computation of $G(p)$ and $\chi_s(p)$ allows us to determine the normalization constants $\calN_1$ and $\calN_2$ since at criticality 
\beq
\begin{split}
G(p) &= \frac{1}{\calN_1^2} \frac{A_d(\Delta_1)}{|p|^{d-2\Delta_1}} , \\ 
\chi_s(p) &= \frac{4}{\calN_2^2} \frac{A_d(\Delta_2)}{|p|^{d-2\Delta_2}} 
\end{split}
\label{ope3} 
\eeq 
for $p\to 0$. The knowledge of $\chi(p_1,p_2,-p_1-p_2)$ then yields the OPE coefficient $c_{112}$ using~\labelcref{ope2}.

\subsection{Effective action} \label{subsec:EA}

The effective action 
\beq 
\Gamma[\phi,h] = - \ln \calZ[J,h] + \int_x \sum_i J_i \phi_i  
\eeq 
is defined as the Legendre transform of the free energy~\cite{Zinn_book}. The source $J$ and the order parameter field $\phi$ are related by 
\beq 
\phi_i(x) = \frac{\delta \ln \calZ[J,h]}{\delta J_i(x)} \quad \mbox{or} \quad J_i(x) = \frac{\delta\Gamma[\phi,h]}{\delta \phi_i(x)} .
\eeq 
All correlation functions for $h=0$ can be obtained from the one-particle irreducible (1PI) vertices 
\begin{multline} 
\Gamma^{(n,m)}_{i_1\cdots i_n}(x_1\cdots x_n,y_1\cdots y_m) = \\  \frac{\delta^{n+m} \Gamma[\phi,h]}{\delta \phi_{i_1}(x_1) \cdots\delta \phi_{i_n}(x_n) \delta h(y_1) \cdots \delta h(y_m) } \biggr|_{\phi=h=0}
\end{multline}
where, assuming the absence of spontaneously broken symmetry, we have set $\phi=0$. In particular, the propagator 
\beq 
G(p) = \Bigl[\Gamma^{(2,0)}_{ii}(p) \Bigr]^{-1} 
\eeq 
is related to the inverse of the two-point vertex computed with a vanishing source $h=0$. The other two correlation functions of interest are given by~\cite{Rose15} 
\beq
\begin{split}
\chi_s(p) &= - \Gamma^{(0,2)}(p) , \\
\chi(p_1,p_2) &= - G(p_1) \Gamma^{(2,1)}_{ii}(p_1,p_2) G(p_2) ,
\end{split}
\eeq 
where we have used the fact that $\Gamma^{(1,1)}_i$ vanishes when evaluated for $\phi=0$. To alleviate the notations we do not write the last argument of the three-point vertices, e.g., $\Gamma^{(2,1)}_{ii}(p_1,p_2)\equiv \Gamma^{(2,1)}_{ii}(p_1,p_2,-p_1-p_2)$.

We are now in a position to relate the OPE coefficient $c_{112}$ to the 1PI vertices at criticality. From Eqs.~(\ref{ope3}) we obtain the normalization constants 
\begin{align}
\calN_1^2 &= A_d(\Delta_1) \lim_{p\to 0}   \frac{\Gamma^{(2,0)}_{ii}(p)}{|p|^{d-2\Delta_1}} , 
\label{N1} \\ 
\calN_2^2 &= - 4 A_d(\Delta_2) \lim_{p\to 0}  \frac{|p|^{2\Delta_2-d}}{\Gamma^{(0,2)}(p)} . 
\label{N2}
\end{align}
Considering~\labelcref{ope2} in the limit $p_2=0$ and $p_1\to 0$, we finally deduce 
\beq 
c_{112} = - \frac{\calN_2}{2\calN_1^2} \frac{A_d(\Delta_1)}{A_d(\Delta_2/2)} \lim_{p\to 0} \frac{\Gamma^{(2,1)}_{ii}(p,0)}{|p|^{\Delta_2-2\Delta_1}} .
\label{c112} 
\eeq 
Equations~\labelcref{N1,N2,c112} are the basic ingredients to determine the OPE coefficient $c_{112}$ in the effective action formalism. In 
\cref{app_free,app_largeN},  
we show that they yield the known results 
\beq 
c_{112} = \sqrt{\frac{2}{N}}
\label{c112free}
\eeq 
for the free theory ($u_0=0$), and 
\beq 
c_{112} = \sqrt{\frac{2}{N}} \left[ \frac{ \Gamma(d-2)}{(d/2-2)} \frac{\sin(\pi d/2)}{\pi} \right]^{1/2} \frac{1}{\Gamma(d/2-1)} 
\label{c112largeN}
\eeq 
 in the large-$N$ limit to the leading order and for $d<4$, in agreement with the literature~\cite{Lang1992a,Lang1994a}.

\subsection{FRG formalism and BMW approximation} 
\label{sec:frgbmw}

The nonperturbative FRG allows one to compute the effective action beyond standard perturbation theory~\cite{Berges02,Delamotte12,Dupuis_review}. Fluctuations are regularized by the infrared regulator term 
\beq 
\Delta S_k[\varphi] = \half \int_q \sum_{i} \varphi_i(-q) R_k(q) \varphi_i(q) , 
\eeq 
where the momentum scale $k$ varies from the UV cutoff $\Lambda$ down to zero. 
A possible choice for the cutoff function $R_k$ is 
\begin{gather}
R_k(q) = Z_k q^2 r\left( \frac{q^2}{k^2} \right),  
\label{eq:def_cutoff}
\end{gather}
with the function $r(y)$ taken to be for instance \begin{gather}
r_\mathrm{W}(y) = \frac{\alpha}{e^y-1}\quad \text{or} \quad
r_\mathrm{E}(y) = \alpha e^{-y}/y.
\end{gather}
$r_\mathrm{W}(y)$ and $r_\mathrm{E}(y)$ define  respectively the so-called Wetterich and exponential regulators. In either case, $\alpha$ is a constant of order one and $Z_k$ a field renormalization factor which varies as $k^{-\eta}$ at criticality~\footnote{We do not introduce a source renormalization factor, see Ref.~\cite{Pagani20} for a FRG approach that introduces explicitly the anomalous dimension of the composite operator.
}.
Thus the regulator suppresses fluctuations with momenta $|q|\lesssim k$ but leaves unaffected those with $|q|\gtrsim k$. The partition function 
\beq 
\calZ_k[J,h] = \int\calD[\varphi] \, e^{-S[\varphi] - \Delta S_k[\varphi] + \int_x (J\varphi+h\varphi^2) }
\eeq 
is now $k$ dependent. The scale-dependent effective action 
\beq
\Gamma_k[\phi,h] = - \ln \calZ_k[J,h] + \int_x \sum_i J_i\phi_i - \Delta S_k[\phi] 
\eeq 
is defined as a slightly modified Legendre transform which includes the subtraction of $\Delta S_k[\phi]$. Assuming that for $k=\Lambda$ the fluctuations are completely frozen by the regulator term,
\beq 
\Gamma_\Lambda[\phi,h] = S[\phi] - \int_x h \phi^2 . 
\eeq 
On the other hand, the effective action of the $\mathrm{O}(N)$ model~\labelcref{action} is given by $\Gamma_{k=0}$ since $R_{k=0}$ vanishes. The FRG approach aims at determining $\Gamma_{k=0}$ from $\Gamma_\Lambda$ using Wetterich's equation~\cite{Wetterich93,Morris94,Ellwanger94}  
\beq 
\dk \Gamma_k[\phi,h] = \half \Tr\llbrace \dk R_k(\Gamma^{(2,0)}_k[\phi,h]+R_k)^{-1} \rrbrace .
\label{eqwet} 
\eeq 
The infinite hierarchy of flow equations satisfied by the $k$-dependent 1PI vertices $\Gamma_k^{(n,m)}$ can be obtained from~\labelcref{eqwet} by taking functional derivatives wrt $\phi$ and $h$.
The presence of the source $h$ in addition to the field $\phi$ allows one to follow the flow of composite fields,
an approach which proved to be useful in tackling a wide range of issues \cite{Pawlowski:2005xe,Igarashi:2009tj,Sonoda:2013dwa,Rose15,Rose:2016elj,Rose:2017lui,Daviet:2018lfy,Pagani:2015hna,Pagani:2016pad,Pagani:2016dof}.

\begin{figure}
	\centerline{\includegraphics[width=6cm]{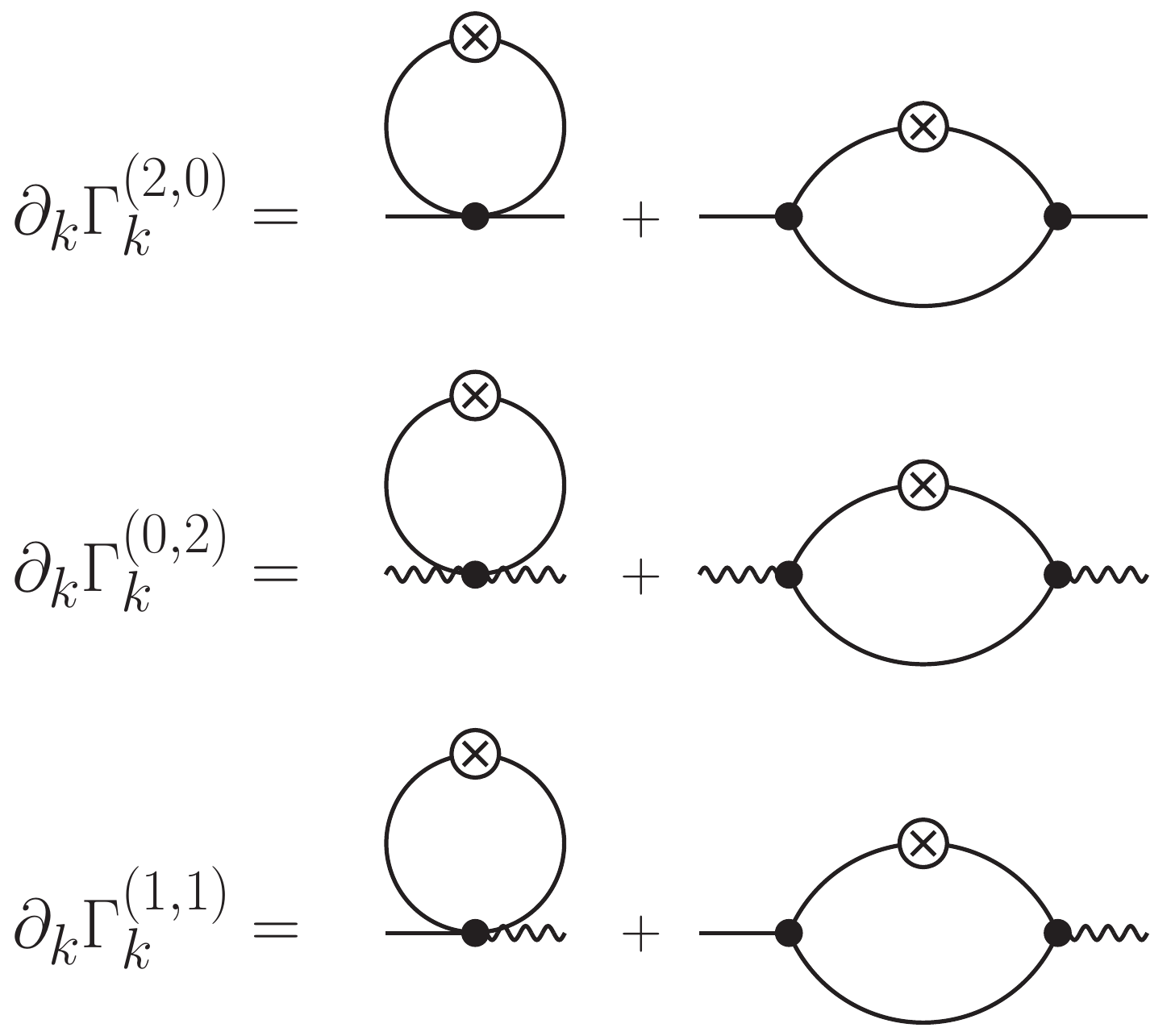}}
	\caption{Diagrammatic representation of the RG equations of $\Gamma^{(2,0)}_k$, $\Gamma^{(0,2)}_k$ and $\Gamma^{(1,1)}_k$. Signs and symmetry factors are not shown. The vertex $\Gamma^{(n,m)}_k$ is represented by a black dot with $n$ solid
		lines and $m$ wavy lines and the solid lines connecting vertices stand for
		the  propagator $G_k = (\Gamma_k^{(2,0)}+R_k)^{-1}$. The cross stands for $\dk R_k$.}
	\label{fig_rgeq_diag} 
\end{figure}

In the BMW approximation~\cite{Blaizot06,Benitez09,Benitez12}, one considers the flow equations of the 1PI vertices in a uniform field $\phi$ even if one is eventually interested in the vanishing field configuration. These equations are shown diagrammatically in \cref{fig_rgeq_diag} for $\Gamma^{(2,0)}_k$, $\Gamma^{(0,2)}_k$ and $\Gamma_k^{(1,1)}$. Since the regulator $\dk R_k$ in \cref{eqwet} restricts the loop momentum to small values $|q|\lesssim k$, whereas the regulator term $\Delta S_k$ ensures that the vertices are regular functions of $p_i^2/k^2$, one can set $q=0$ in the vertices $\Gamma_k^{(n,m)}$. Noting then that a vertex with a vanishing momentum can be related to a lower-order vertex, e.g.,
\beq 
\begin{split}
\Gamma^{(3,0)}_{k,ijl}(\p,-\p,0;\phi) &= \frac{\partial\Gamma^{(2,0)}_{k,ij}(\p;\phi)}{\partial \phi_l} , \\ 
\Gamma^{(2,1)}_{k,ij}(\p,0,-\p;\phi) &= \frac{\partial\Gamma^{(1,1)}_{k,i}(\p;\phi)}{\partial \phi_j} , 
\end{split}
\label{Gamma21}
\eeq 
we obtain a closed set of equations satisfied by $\Gamma_k^{(2,0)}(\p,\phi)$, $\Gamma_k^{(0,2)}(\p,\phi)$ and $\Gamma_k^{(1,1)}(\p,\phi)$; see Ref.~\cite{Rose15} for the explicit expressions. These equations, together with the expression~\labelcref{Gamma21} of $\Gamma^{(2,1)}_{k}(\p,0,-\p;\phi)$ are sufficient to obtain the vertices necessary to determine the normalization constants $\calN_1$, $\calN_2$ and the OPE coefficient $c_{112}$. 

The reader may wonder if the simple choice of bare operators
made in (\ref{def:composite-op-initial-cond}) is suitable to address our objectives. Thus, let us expound on some fundamental properties of composite operators within the FRG formalism.
In the FRG framework a composite operator can be defined by differentiating the effective action $\Gamma_k$ wrt an external source.
For instance, the running composite operator corresponding to $\calO_2$ is defined as
$\left[\phi^2\right]\left(k\right)\equiv -\frac{\delta \Gamma_k}{\delta h}[\phi,h]|_{h=0}$.
$\left[\phi^2\right]\left(k\right)$ depends on the RG scale $k$ and, if evaluated at
the UV cutoff scale, satisfies $\left[{\phi^2} \right]\left(\Lambda \right)=\phi^2$.
As soon as one lowers the scale $k$ from $\Lambda$, the bare term gets renormalized through its coupling to the 1PI vertices. This implies that the flow of the composite operator generates mixing with other operators in the sense that $\left[\phi^2 \right]\left(k \right) = Z_{22,k}\phi^2+Z_{24,k}\phi^4+\cdots $.

It must be noted, however, that scaling operators are not just mere composite operators. A scaling operator is a particular combination of composite operators that diagonalize the flow linearized about the fixed point.
As a consequence,  in the scaling regime, $\left[\phi^2 \right]\left(k \right)$  can be expressed as a linear combination of scaling operators. 
However, by lowering the RG scale $k$ to $0$ in such a linear combination,
the scaling operators of higher scaling dimension $\Delta_i$
are suppressed and only the lowest scaling operator survives.
Indeed, numerically solving the flow of e.g. $\Gamma_k^{(0,2)}$ within our approximation scheme, we have been able to check
that our procedure reproduces the expected behavior of a scaling operator in the fixed point
regime~\footnote{This reasoning could not be applied if one were interested in some other scaling operator, say the one having scaling dimension $\Delta_4$.}.

Let us conclude by noticing that the dependence of the vertices
$\Gamma_k^{(2,0)}$, $\Gamma_k^{(0,2)}$, and $\Gamma_k^{(1,1)}$
on the background field $\phi$ entails an infinite number of 1PI vertices (albeit in a specific momentum configuration).
This shows that our ansatz includes non-trivial mixing among field monomials $\phi^2$, $\phi^4$ and so on.

\section{Numerical results} 
\label{sec_results} 

The flow equations are integrated numerically, see e.g. Refs.~\cite{Rose15,Benitez12} for details. We work with dimensionless variables, $\tilde p = p/k$  and $\tilde \rho = Z_k k^{2-d} \rho$. The field dependence of the potential and the vertices is discretized on a finite and evenly spaced grid $\tilde \rho \in [0,\tilde\rho_\text{max}]$ comprising  $N_\rho$ points, while the momentum dependence of the vertices is approximated by Chebyschev polynomials of order $N_p$ defined on $[0,\tilde p_\text{max}]$. The integration of the flow with respect to the RG scale $k$ is done with an adaptive step integration. Convergence of the results with respect to the parameters has been verified; their typical range are  $N_\rho=40$--$80$, $\tilde\rho_\text{max}=4$--$8$, $\tilde p_\text{max}=4$--$10$ and $N_p=20$--$30$ with the precise value depending on $d$ and $N$.

For each universality class set by $d$ and $N$ and each choice of the cutoff function \labelcref{eq:def_cutoff} parameterized by $\alpha$, the critical point is found by tuning the initial condition of the flow. This enables the computation of $G(p)$, $\chi_s(p)$ and $\Gamma^{(2,1)}(p,0)$  [\cref{ope3,Gamma21}] at criticality, from which one fits the values of the critical exponents $\eta$ and $\nu$ (or equivalently $\Delta_{1,2}$) and normalization constants $\mathcal{N}_{1,2}$, yielding $c_{112}$ through \cref{c112}.

A crucial question is that of the regulator dependence. Indeed, while \cref{eqwet} is exact, any approximation scheme such as BMW introduces a regulator dependence to the results. 
In order to provide a meaningful prediction for a physical quantity $Q(\alpha)$, a choice of the regulator must be made. The usual rationale is the so-called principle of minimum sensitivity (PMS), according to which the best value of $\alpha$ is that for which the regulator dependence of $Q(\alpha)$ is minimal, i.e., for which $\partial_\alpha Q(\alpha)=0$, or failing that for which $|\partial_\alpha Q(\alpha)|$ is minimal.

\begin{figure}
	\centerline{\includegraphics{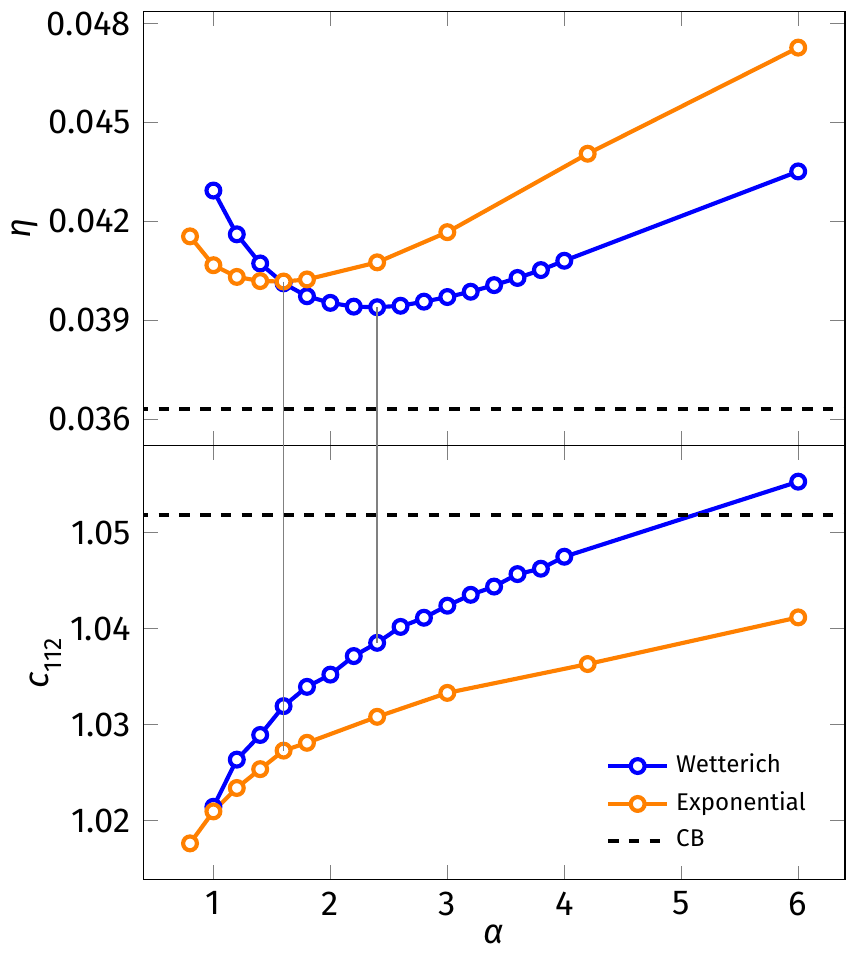}}
	\caption{Dependence of the critical exponent $\eta$ (top) and the OPE coefficient $c_{112}$ (bottom) of the $3d$ Ising universality class ($N=1$) on the coefficient $\alpha$ for the Wetterich (blue) and exponential (orange) regulators. For reference, the conformal bootstrap estimates $\eta_{\mathrm{CB}}=0.036308$ and $c_{112,\mathrm{CB}}=1.0518537$ are shown as black dashed lines~\cite{Kos16}. Applying the PMS yields optimal parameters $\alpha_\mathrm{E}=1.6$ and $\alpha_\mathrm{W}=2.4$ for the exponential and Wetterich regulators (respectively subscript E and W), shown by gray vertical lines, with corresponding values $\eta_{\mathrm{E}}=0.0402$, $c_{112,\mathrm{E}}=1.027$, $\eta_{\mathrm{W}}=0.0394$ and $c_{112,\mathrm{W}}=1.039$. The value we retain for $c_{112}$ is $c_{112,\mathrm{W}}$, given that it corresponds to the extremal value of $\eta$ across both families of regulators. 
		}
	\label{fig_PMS} 
\end{figure}

However, the PMS  for $c_{112}$, shown for the $3d$ Ising universality ($N=1$) class in \cref{fig_PMS}, does not provide a satisfactory result.  Indeed, for a given regulator, $c_{112}$ is a monotonous concave function of $\alpha$, with no extremum or inflection point, varying by about $3\%$ over the range of regulators considered. As a consequence, we choose the regulator that fulfills the PMS for the anomalous dimension $\eta$. The value thus obtained for $c_{112}$ depends only weakly on the family of regulators considered, with a variation of about $1.15\%$ between the Wetterich and exponential regulators. The regulator dependency is slightly smaller than the $1.2\%$ difference with the conformal bootstrap estimate.

As a side note, we  point out a recent proposal for an alternative way to fix the regulator dependence for conformally invariant theories, the principle of maximal conformality (PMC)~\cite{Balog2020a}. Conformal invariance implies a set of (modified) Ward identities associated with scale and special conformal transformations (SCT). While scale invariance is always fulfilled at the fixed point,  invariance under SCT is broken within the derivative expansion at high order. PMC suggests to choose the regulator that minimizes the symmetry breaking. 
While in the present case it is not straightforward to implement the PMC for BMW, because the Ward identities are either trivially fulfilled or involve high-order vertices that cannot be computed using the BMW approximation, its implementation  for  the derivative expansion shows that the PMS for $\eta$ and PMC yield very close results, 
providing a further argument in favor of our regulator choice.  

\subsection{Ising university class in dimensions $2\leq d\leq 4$}

\begin{figure}
	\centerline{\includegraphics[]{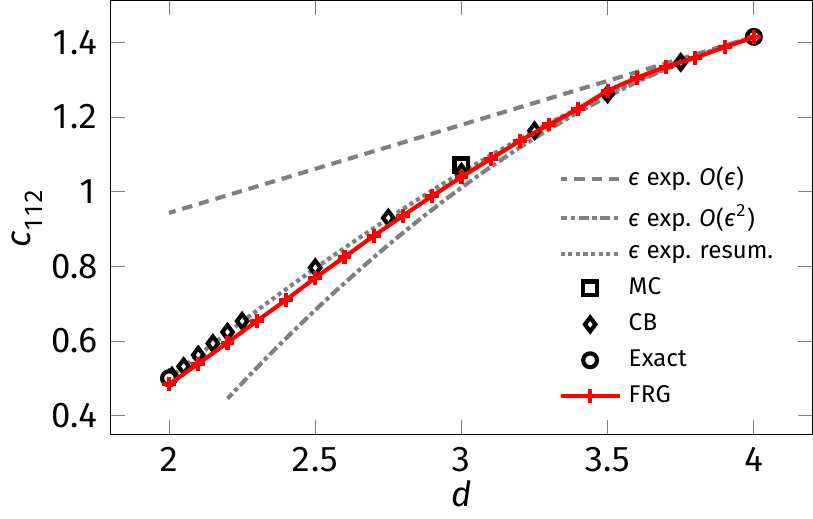}}
	\caption{OPE coefficient $c_{112}$ of the Ising university class as a function of the dimension $d$. The solid red crosses are obtained from FRG, with the full line a guide to the eye. The black symbols correspond to reference estimates from  Monte Carlo (square)~\cite{Caselle2015a} and conformal bootstrap (diamond)~\cite{Kos16,Cappelli2019a} and the exact values in $d=2$ and $4$. The gray lines are given by the $\epsilon$~expansion [\cref{eq:c112epsIsing}] about $d=4$ to order $\calO(\epsilon)$ (dashed),  $\calO(\epsilon^2)$ (dashdotted) and the $(3,2)$ resummation imposing the $2d$ result (dotted)~\cite{Carmi2021a}.}
	\label{fig_c112vsD} 
\end{figure}

We first consider the OPE coefficients of the Ising university class ($N=1$) for dimensions $d$ between the lower and upper critical dimensions $d=2$ and $4$, for which the results are shown in \cref{fig_c112vsD} and \cref{tab:c112vsD}. The FRG results can be compared to the exact values in $d=2$ and $4$, the conformal bootstrap~\cite{Kos16} and Monte Carlo~\cite{Caselle2015a} estimates in $d=3$ and the $\epsilon$ expansion up to  fourth order~\cite{Gopakumar2017a,Gopakumar2017b,Dey2017a,Carmi2021a},
\begin{align}
    c_{112}={}&\sqrt{2}\bigg(1-\frac{1 }{6}\epsilon-\frac{77 }{648}\epsilon ^2\nonumber\\
    &{}+\frac{3726 \zeta (3)-1915 }{34992}\epsilon ^3+A_4 \epsilon^4\bigg) + \mathcal{O}(\epsilon^5)
\label{eq:c112epsIsing}
\end{align}
where $\zeta (3)$ is  Ap\'{e}ry's  constant and $A_4=- 0.158947\ldots$ is only known numerically~\cite{Carmi2021a}.

Owing to its asymptotic nature, the $\eps$ expansion does not converge; indeed, for $d=3$, its relative error to all estimates (FRG, conformal bootstrap, Monte Carlo) increases from $4\%$ at order $\calO(\epsilon^2)$ to $6\%$ at order $\calO(\epsilon^3)$ and $16\%$ at order $\calO(\epsilon^4)$. In order to make sense of the results, a resummation procedure must be carried out, for instance by approximating $c_{112}(\eps)$ by a Pad\'e approximant of order $(n,m)$, i.e. a rational fraction of $\eps$ with numerator and denominator of degree $n$ and $m$, respectively. Following Ref.~\cite{Carmi2021a}, we pick an approximant of order $(3,2)$, whose coefficients are uniquely determined by imposing the $\calO(\eps^4)$ expansion  around  $\eps=0$ and the exact value $c_{112}(\eps=2)=1/2$. This gives $c_{112}=1.0507$ for $d=3$, in very good agreement with the conformal bootstrap, to be compared to the $0.8\%$ error when the approximant of order $(3,1)$ is used and the exact result for $d=2$ is not imposed. 
Let us mention that in this case different choices of Pad{\'e} approximants lead to somewhat different results, which may be used to estimate an average value and its uncertainty: $c_{112}=1.048(31)$, see Table~\ref{tab:c112vsD}

\begin{table}[t]
\centering
\begin{tabular}{lS[table-format=1.4]S[table-format=1.11]S[table-format=1.3]}
\hline
\hline
\multicolumn{1}{c}{$d$} & \multicolumn{1}{c}{$2$} & \multicolumn{1}{c}{$3$} & \multicolumn{1}{c}{$4$}\\
\hline 
{FRG}    & 0.484 & 1.039                         & 1.413\\
 $\eps$ exp. $(3,1)$ &  0.4259   & 1.0432 & {$\sqrt{2}\simeq 1.414$}\\
 $\eps$ exp. $(2,2)$ & -0.0698   & 1.0200 & {$\sqrt{2}\simeq 1.414$}\\
 $\eps$ exp. $(1,3)$ &  0.7442   & 1.0805 & {$\sqrt{2}\simeq 1.414$}\\
 $\eps$ exp.${}+2d$ (3,2)          &   & 1.0507 & {$\sqrt{2}\simeq 1.414$}\\
 $\eps$ exp.${}+2d$ (2,3)          &   & 1.0464 & {$\sqrt{2}\simeq 1.414$}\\
MC~\hfill\cite{Caselle2015a}     &        & 1.07(3) &      \\
CB~\hfill\cite{Kos16}    & & 1.0518537(41) & \\  
Exact~\hfill\cite{Francesco1997a} & {$1/2=0.5$} & & {$\sqrt{2}\simeq 1.414$} 
 \\
\hline
\hline
\end{tabular}
\caption{OPE coefficient $c_{112}$ of the Ising universality class for dimensions $d=2$, $3$, $4$. 
We compare the numerical FRG results to:
(i) various resummations of the $\epsilon$~expansion [\cref{eq:c112epsIsing}] to order $\calO(\eps^4)$
obtained by employing different Pad{\'e} approximants and by possibly
taking into account the exactly known 2$d$ value,
(ii) conformal bootstrap and (iii) Monte Carlo estimates and (iv) the exact values for the $2d$ and $4d$ Ising universality classes.
\label{tab:c112vsD}}
\end{table}

Compared to the best results (exact in $d=2$ and $4$, conformal bootstrap and Monte Carlo in $2<d<4$), the FRG always has an error smaller than $3\%$, with less than $2\%$ error in $d=3$.  
By contrast with the  resummed  $\eps$ expansion, which requires   additional  input in the form of the $2d$ data to provide accurate results, the FRG and conformal bootstrap are able to interpolate smoothly between dimensions $d=2$ and $4$. As the dimension is increased, $c_{112}$ increases monotonously, with an almost linear behavior between $d=2$ and $d=3$, which might partly explain the remarkable agreement of the $\epsilon$ expansion resummation (when supplemented with the exact $2d$ result) with conformal bootstrap, Monte Carlo and FRG.
However, if the resummation is not supplemented with the $2d$ result
then the $\epsilon$-expansion estimate of $c_{112}$ is rather poor close
to and at $d=2$.

In $d=4$, the FRG within the BMW approximation scheme gives the exact analytic value of $c_{112}$. The small difference ($\sim 0.1\%$) between the numerical result and the exact value seen in~\cref{tab:c112vsD} arises from the fitting of the critical exponents and the normalization constants. This serves as an estimate of this numerical error: in lower dimensions, it is much smaller than the difference to the best estimates.

\subsection{Three-dimensional $\mathrm{O}(N)$ model} 

\begin{figure}
	\centerline{\includegraphics[]{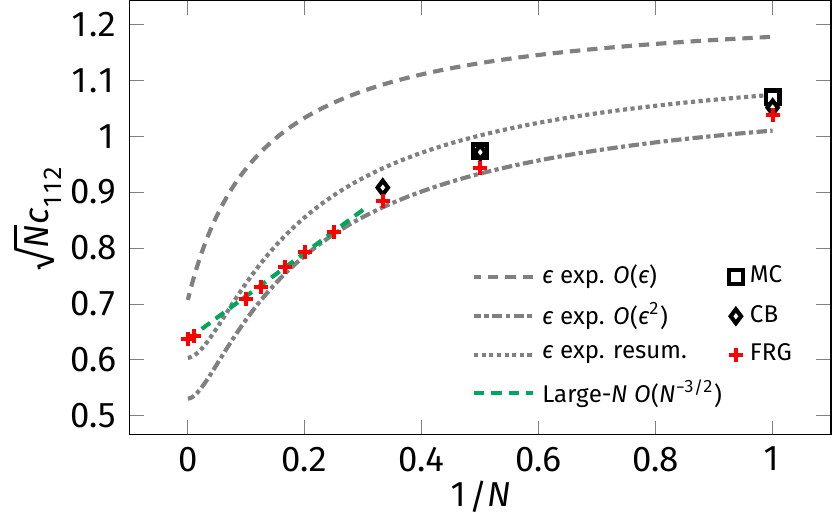}}
	\caption{Rescaled OPE coefficient $\sqrt{N}c_{112}$ of the three-dimensional $\mathrm{O}(N)$ model as a function of the inverse number of field components $1/N$. The solid red crosses are obtained from FRG. The horizontal dashed green line shows the large-$N$ result to leading order. The gray lines are given by the order $\calO(\epsilon)$ (dashed), 
	$\calO(\epsilon^2)$ (dashdotted) and the $(2,1)$ resummation (dotted) of the  $\epsilon$~expansion~\cite{Dey2017a}. The black symbols correspond to  estimates from  Monte Carlo (square)~\cite{Caselle2015a,Hasenbusch2020a} and conformal bootstrap (diamond)~\cite{Kos16,Chester20,Chester2020a}. 
	}
	\label{fig_c112vsN} 
\end{figure}

We now focus on the three-dimensional $\mathrm{O}(N)$ model. Given that the large-$N$ result is~\cite{Lang1992a,Lang1994a,Alday:2019clp} 
\begin{equation}
    c_{112}  = \dfrac{2}{\pi} \dfrac{1}{\sqrt{N}}+ \dfrac{24}{\pi^3}\dfrac{1}{N^{3/2}}+\calO\bigg(\dfrac{1}{N^{5/2}}\bigg),
\end{equation}
we consider rather than $c_{112}$ the rescaled OPE coefficient $\sqrt{N}c_{112}$ that has a well-defined large-$N$ limit. FRG results and estimates from the $\epsilon$ expansion~\cite{Dey2017a}, conformal bootstrap~\cite{Kos16,Chester20,Chester2020a} and Monte Carlo~\cite{Caselle2015a,Hasenbusch2020a} are shown in \cref{fig_c112vsN} and \cref{tab:c112vsN}. For $N>1$, $c_{112}$ is only known up to order $\calO(\eps^3)$ and  we resum the $\eps$~expansion using a $(2,1)$ Pad\'e approximant.

\begin{table}[b]
\centering
\begin{tabular}{lllll}
\hline
\hline
\multicolumn{1}{c}{$N$} & \multicolumn{1}{c}{FRG} & \multicolumn{1}{c}{$\epsilon$ exp.} & \multicolumn{1}{c}{MC} &  \multicolumn{1}{c}{CB}\\
\hline
$1$    & $1.039$ & $1.075$~\cite{Dey2017a} & $1.07(3)$\hfill~\cite{Caselle2015a} & $1.0518537(41)$\hfill~\cite{Kos16} \\
$2$    & $0.943$ & $1.002$~\cite{Dey2017a} & $0.9731(14)$\hfill~\cite{Hasenbusch2020a}& $0.971743(38)$\hfill~\cite{Chester20}    \\
$3$    & $0.884$ & $0.943$~\cite{Dey2017a}&       & $0.908047(102)$\hfill~\cite{Chester2020a}\\
$4$    & $0.829$ & $0.895$~\cite{Dey2017a}& \\
$5$    & $0.792$ & $0.855$~\cite{Dey2017a}& \\
$6$    & $0.766$ & $0.823$~\cite{Dey2017a}& \\
$8$    & $0.731$ & $0.773$~\cite{Dey2017a}& \\
$10$   & $0.709$ & $0.737$~\cite{Dey2017a}& \\
$100$  & $0.643$ & $0.607$~\cite{Dey2017a}& \\
$1000$ & $0.638$ & $0.603$~\cite{Dey2017a}& \\
\hline
\hline
\end{tabular}
\caption{Rescaled OPE coefficient $\sqrt{N}c_{112}$ of the three-dimensional $\mathrm{O}(N)$ model for different numbers of field components $N$. We compare the FRG results to the  $(2,1)$ resummation of the $\epsilon$ expansion to order $\calO(\epsilon^3)$,  conformal bootstrap and Monte Carlo estimates. The exact large-$N$ result is  $\lim_{N\to\infty}\sqrt{N}c_{112} =2/\pi\simeq 0.637$.
\label{tab:c112vsN}}
\end{table}

For  $N=1$, $2$, $3$, FRG differs from conformal bootstrap by respectively $1.2\%$, $3.0\%$ and $2.6\%$. Furthermore FRG accurately reproduces the large-$N$ behavior: for $N=1000$, the FRG estimate $\sqrt{N}c_{112} = 0.638$ differs from the exact large-$N$ result $\lim_{N\to\infty}\sqrt{N}c_{112} =2/\pi\simeq 0.637$ by $0.1\%$,
 which is about the order of magnitude corresponding to a 
$1/N$ correction.
This is expected as it is known that the relevant vertices are exact in the large-$N$ limit~\cite{Blaizot06,Rose15}.
By contrast, the $(2,1)$ resummation of the $\epsilon$~expansion gives 
$\lim_{N\to\infty}\sqrt{N}c_{112}=\sqrt{2}(4 \zeta (3)+1)/(8 \zeta (3)+4)\simeq 0.603$ with a $5.3\%$ error.

Moreover, numerically fitting the FRG data for $N\gtrsim 5$ by a law of the form
$
    c_{112}  = 2/\pi\sqrt{N} + \kappa/N^{3/2}
$
yields $\kappa\simeq 0.76$, in very good agreement with the exact value $24/\pi^3\simeq 0.774$.

\section{Conclusion}

We have shown how to extract the OPE coefficients of a conformal theory within the framework of FRG, by determining three-point vertices in specific momentum configurations. We have used our approach to determine the $c_{112}$ coefficient, corresponding to the simplest possible OPE coefficient, in the $\mathrm{O}(N)$ universality class for various $d$ and $N$. This provides the first
non-perturbative determination of the OPE coefficients based on field theory, aside from 
the lattice computations in \cite{Caselle2015a,Hasenbusch2020a} and 
the conceptually  different conformal bootstrap.

While the  accuracy of FRG can be sometimes difficult to gauge in the absence of a small expansion parameter, the fact that
the results compare extremely well with the values, when available, obtained from Monte Carlo and conformal bootstrap increases confidence in the validity of the method.
 It is a testament to the versatility of FRG that, in this specific case, tuning such parameters as $d$ or $N$  demands relatively little effort as they  only enter the flow equations through their explicit values.

Lastly, we note that the OPE can be used in settings very different from the critical $\mathrm{O}(N)$ models investigated in this work, for instance in theories away from a fixed point or at a non-equilibrium fixed point. In these cases many methods holding for equilibrium critical theories are not available. Our work suggests that the FRG may constitute the right framework to tackle these issues thanks to its aforementioned versatility and to the fact that the FRG equations can be solved without invoking further requirements, such as conformal symmetry.

\begin{acknowledgments}
The authors thank Gonzalo~De~Polsi, Matthieu~Tissier and Nicolás~Wschebor for stimulating discussions and are extremely grateful towards Johan~Henriksson
 for pointing out Refs.~\cite{Lang1994a,Cappelli2019a,Carmi2021a}. C.~P.~and N.~D.~wish to thank the organizers of the meeting FRGIM 2019 where some of the ideas developed in this work were discussed.
C.~P.~thanks Hidenori~Sonoda for discussions and collaboration on closely related projects
and the Laboratoire de Physique Th{\'e}orique de la Mati{\`e}re Condens{\'e}ee in Paris for hospitality.

C.~P.~acknowledges support by the DFG grant PA 3040/3-1.
F.~R. acknowledges support from the Deutsche Forschungsgemeinschaft (DFG, German Research Foundation) under Germany’s Excellence Strategy--EXC--2111--390814868.
\end{acknowledgments}

\appendix

\section{$c_{112}$ in the free case} 
\label{app_free} 

When $u_0$ vanishes, the functional integral over the field $\varphi$ can be done exactly and yields the partition function 
\beq 
Z[J,h] = e^{\frac{N}{2} \Tr\ln G[h] + \half \int_{x,y} \sum_i J_i(x) G[x,y;h] J_i(y) } ,
\eeq 
where $G[h]$ denotes the propagator in the presence of an arbitrary external source $h$: 
\beq 
G^{-1}[x,y;h] = \bigl(-\nabla^2_x + 2 h(x)\bigr) \delta(x-y) . 
\eeq 
The expectation value of the field is given by 
\beq 
\phi_i(x) = \int_y G[x,y;h] J_i(y) , 
\eeq 
and the effective action is simply 
\beq 
\Gamma[\phi,h] = \int_x \llbrace \half (\dmu\phi)^2 - h\phi^2 \rrbrace - \frac{N}{2} \Tr\ln G[h] . 
\eeq 
We thus obtain 
\begin{align}
\Gamma^{(2,0)}(p) &= p^2 , \label{app1} \\ 
\Gamma^{(0,2)}(p) &= -2N \int_q \frac{1}{ q^2 (p+q)^2} \simeq -  \frac{2NB_d}{|p|^{4-d}} , \label{app2} \\ 
\Gamma^{(2,1)}(p_1,p_2) &= -2 \label{app3} ,
\end{align} 
where 
\beq 
B_d = \frac{1}{(4\pi)^{d/2}} \frac{\Gamma(2-d/2) \Gamma(d/2-1)^2 }{\Gamma(d-2) } .
\label{Bd} 
\eeq 
The last expression for $\Gamma^{(0,2)}(p)$ is obtained for $p\to 0$ and $d<4$. From~(\ref{app1}) and (\ref{N1}) we deduce $\Delta_1=d/2-1$  (in agreement with the vanishing anomalous dimension in the free theory) and 
\beq 
\calN_1^2=A_d(d/2-1) . 
\eeq 
On the other hand, comparing~(\ref{app2}) with (\ref{N2}) yields $\Delta_2=d-2$ (in agreement with $\nu=1/2$) and 
\begin{align}
\frac{1}{\calN_2^2} &=  \frac{N}{2} \frac{B_d}{A_d(d-2)}  \nonumber\\ 
&= \frac{N}{32\pi^d} \Gamma(d/2-1)^2 . 
\end{align} 
From~(\ref{c112}) and (\ref{app3}) we finally obtain~(\ref{c112free}).

\section{$c_{112}$ in the large-$N$ limit} 
\label{app_largeN} 

\def\Tkt{T_{\rm BKT}}
\def\bdelta{\bar\delta} 
\def\GamA{\Gamma_{A,k}}
\def\GamB{\Gamma_{B,k}}
\def\GamBL{\Gamma_{B,\Lambda}}
\def\DA{\Delta_{A,k}}
\def\DB{\Delta_{B,k}}
\def\YA{Y_{A,k}}
\def\YB{Y_{B,k}}

Following the Appendix of Ref.~\cite{Rose15}, we introduce the field $\rho=\varphi^2$ and a Lagrange multiplier $\lambda$ to write the partition function $\calZ[h]\equiv \calZ[J=0,h]$ of the $\mathrm{O}(N)$ model as 
\begin{align}
	\calZ[h] ={}& \int \calD[\varphi,\rho,\lambda] \exp \biggl\lbrace -\int_x \Bigl[ \half(\nablabf\varphi)^2 \nonumber\\ & + \left(\frac{r_0}{2}-h\right) \rho  +  
	\frac{u_0}{4!N} \rho^2 + i \frac{\lambda}{2} (\varphi^2-\rho) \Bigr] 
	\biggr\rbrace \nonumber \\ 
	={}& \int \calD[\varphi,\lambda] \exp\biggl\lbrace \int_x \biggl[ \frac{3N}{2u_0} (2h+i\lambda-r_0)^2 \nonumber\\ & 
	-\half \left[ (\nabla\varphi)^2 + i \lambda \varphi^2 \right] 
	\biggr\rbrace .
\end{align}
Then we split the field $\varphi$ into a $\sig$ field and an $(N-1)$-component field $\pi$. Integrating over the $\pi$ field, we obtain the action 
\begin{multline}
	S[\sig,\lamb,h] = \int_x \biggl[ -\frac{3N}{2u_0} (2h+i\lambda-r_0)^2 \\
	+ \half \left[ (\nabla\sig)^2 + i \lambda \sig^2 \right] 
	+ \frac{N-1}{2} \Tr \ln g^{-1}[\lamb] , 
\end{multline} 
where 
\beq 
g^{-1}[x,x';\lamb] = [-\nabla_x^2 + i\lamb(x) ] \delta(x-x') 
\eeq
is the inverse propagator of the field $\pi_i$ in the fluctuating $\lamb$ field. In the limit $N\to\infty$, the action becomes proportional to $N$ (if one rescales the $\sig$ field, $\sig\to\sqrt{N}\sig$); the saddle point approximation becomes exact for the partition function $\calZ[h]$ and the Legendre transform of the free energy coincides with the action $S$~\cite{Lebellac_book}. This implies that the effective action is simply equal to $S[\sig,\lamb,h]$:
\begin{multline}
	\Gamma[\sig,\lamb,h] = \int_x \biggl\lbrace -\frac{3N}{2u_0} (2h+i\lambda-r_0)^2 \\
	+ \half \left[ (\nabla\sig)^2 + i \lambda \sig^2 \right] \biggr\rbrace 
	+ \frac{N}{2} \Tr \ln g^{-1}[\lamb] 
\label{app4}
\end{multline} 
(we use $N-1\simeq N$ for large $N$). We can eliminate the Lagrange multiplier $\lamb$ using 
\beq 
\frac{\delta \Gamma[\sig,\lamb,h]}{\delta \lamb(x)}\biggl|_{\lamb=\lamb[\sig,h]} = 0  
\eeq 
to obtain the effective action $\Gamma[\sig,h]\equiv \Gamma[\sig,\lamb[\sig,h],h]$, which is the starting point to compute the vertices $\Gamma^{(n,m)}$ in the large-$N$ limit.

In Ref.~\cite{Rose15} it was shown that, at criticality, 
\beq 
\begin{split}
\Gamma^{(2,0)}(p) &= p^2 , \\ 
\Gamma^{(0,2)}(p) &= - \frac{12N}{u_0} + \left(\frac{6N}{u_0} \right)^2 \Gamma^{(2)}_{\lamb\lamb}(p)^{-1}  , 
\end{split}
\eeq 
where
\beq 
\begin{split}
\Gamma^{(2)}_{\lamb\lamb}(p) &= \frac{3N}{u_0} + \frac{N}{2} \Pi(p) , \\ 
\Pi(q) &= \int_q \frac{1}{q^2(p+q)^2} \simeq  \frac{B_d}{|p|^{4-d}}  
\end{split}
\eeq 
for $p\to 0$ and $d<4$, where $B_d$ is defined in~(\ref{Bd}). Calculating the three-point function along the same lines, one finds 
\begin{align}
\Gamma^{(2,1)}(p_1,p_2) &= - \frac{6N}{u_0} \Gamma^{(2)}_{\lamb\lamb}(p_1+p_2)^{-1} \nonumber\\ 
&\simeq - \frac{12 |p_1+p_2|^{4-d}}{u_0 B_d} 
\end{align} 
for $p_1+p_2\to 0$ and $d<4$. From~(\ref{N1}) and (\ref{N2}) one then obtains $\Delta_1=d/2-1$ and $\Delta_2=2$ (in agreement with the large-$N$ results $\eta=0$ and $\nu=1/(d-2)$ to leading order) and 
\beq 
\begin{split}
\calN_1^2 &= A_d(d/2-1) , \\ 
\calN_2^2 &= - \frac{u_0^2 B_d A_d(2)}{18N} . 
\end{split}
\eeq
Equation~(\ref{c112}) then gives
\beq 
c_{112} = \left[ - \frac{2A_d(2)}{NB_d} \right]^{1/2} \frac{1}{A_d(1)} 
\eeq 
and in turn~(\ref{c112largeN}) using standard properties of the Gamma function.

\end{document}